\documentclass[aps,prl,twocolumn,superscriptaddress,showpacs]{revtex4-1}
\makeatletter

\newcommand{\Rmnum}[1]{\expandafter\@slowromancap\romannumeral #1@}
\makeatother
\usepackage{graphicx}
\usepackage{latexsym}
\usepackage{amssymb}
\usepackage{amsfonts}
\usepackage{soul}
\usepackage{color}
\usepackage{amsmath}
\usepackage{natbib}
\usepackage[english]{babel}

\begin{document}

\title{Direct visualization of vortex ice in a nanostructured superconductor}

\author{Jun-Yi Ge}\email{Junyi.Ge@kuleuven.be}
\affiliation{INPAC--Institute for Nanoscale Physics and Chemistry, KU Leuven, Celestijnenlaan 200D, B--3001 Leuven, Belgium}
\author {Vladimir N. Gladilin}
\affiliation{INPAC--Institute for Nanoscale Physics and Chemistry, KU Leuven, Celestijnenlaan 200D, B--3001 Leuven, Belgium}
\affiliation{TQC--Theory of Quantum and Complex Systems, Universiteit Antwerpen, Universiteitsplein 1, B--2610 Antwerpen, Belgium}
\author{Jacques Tempere}
\affiliation{TQC--Theory of Quantum and Complex Systems, Universiteit Antwerpen, Universiteitsplein 1, B--2610 Antwerpen, Belgium}
\author{Vyacheslav S. Zharinov}
\affiliation{INPAC--Institute for Nanoscale Physics and Chemistry, KU Leuven, Celestijnenlaan 200D, B--3001 Leuven, Belgium}
\author{Joris Van de Vondel}
\affiliation{INPAC--Institute for Nanoscale Physics and Chemistry, KU Leuven, Celestijnenlaan 200D, B--3001 Leuven, Belgium}
\author{Jozef T. Devreese}
\affiliation{TQC--Theory of Quantum and Complex Systems, Universiteit Antwerpen, Universiteitsplein 1, B--2610 Antwerpen, Belgium}
\author{Victor V. Moshchalkov}
\affiliation{INPAC--Institute for Nanoscale Physics and Chemistry, KU Leuven, Celestijnenlaan 200D, B--3001 Leuven, Belgium}

\date{\today}

\begin{abstract}
Artificial ice systems have unique physical properties promising for potential applications. One of the most challenging issues in this field is to find novel ice systems that allows a precise control over the geometries and many-body interactions. Superconducting vortex matter has been proposed as a very suitable candidate to study artificial ice, mainly due to availability of tunable vortex-vortex interactions and the possibility to fabricate a variety of nanoscale pinning potential geometries. So far, a detailed imaging of the local configurations in a vortex-based artificial ice system is still lacking. Here we present a direct visualization of the vortex ice state in a nanostructured superconductor. By using the scanning Hall probe microscopy, a large area with the vortex ice ground state configuration has been detected, which confirms the recent theoretical predictions for this new ice system. Besides the defects analogous to artificial spin ice systems, other types of defects have been visualized and identified. We also demonstrate the possibility to realize different types of defects by varying the magnetic field.
\end{abstract}

\maketitle

\section {1. Introduction}
Spin ordering in condensed matter materials crucially affects their fundamental properties.
Frustration occurs when the energy of interaction between adjacent spins cannot be simultaneously minimized \cite{Wannier}. Examples include the rare-earth oxide pyrochlores where the rare-earth ions form a lattice of corner-sharing tetrahedra \cite{Harris,Siddharthan,Ramirez,Bramwell,Petit}. In such materials, the minimization of interaction energy requires the tetrahedron spins to follow the "two-in, two-out" Pauling ice rule which determines the proton positional ordering in the archetype water ice \cite{Pauling}. Thus it is named ``spin ice''. Spin ice materials exhibit a variety of intriguing phenomena, such as residual entropy in the low-temperature limit \cite{Ramirez,Lau,Pomaranski}, massively degenerate ground states \cite{Perrin} and emergent excitations of magnetic monopoles and Dirac strings \cite{Castelnovo,Morris,Ladak,Morgan,Tokiwa}. Besides the fundamental interest in understanding the exotic phases in magnetism, the spin ice materials also have great potential in spintronics applications, such as energy storage, memory storage and logic devices \cite{Vedmedenko,Heyderman,Ortiz,Wang2016}.

However, the difficulty of studying spin ice materials arises from the fact that probing individual spins without altering the state of the system is quite challenging. It is also difficult to tune the lattice spacing/geometry as well as to locally control the defects. To avoid this problem, single-domain ferromagnetic islands have been lithographically patterned into various frustrated geometries which mimic spin ice materials \cite{Wang,Moller,Qi,Zhang,Farhan,Gilbert,Gilbert-2,Zhou}.  
These patterned ferromagnetic islands, known as artificial spin ice (ASI), enable direct probing of local configurations in ice systems.
Moreover, designing and functionalizing the complex interactions in ice systems become accessible in ASI systems. 
However, the relatively weak dipole interactions between ferromagnetic islands make it difficult to reach the long-range ordered ground state in ASI systems \cite{Ke,Li,Budrikis}. Finding new systems which behave as analogues of ice systems could be an important extension to this interesting field of research. 

\begin{figure*}[!t]
\centering
\includegraphics*[width=0.9\linewidth,angle=0]{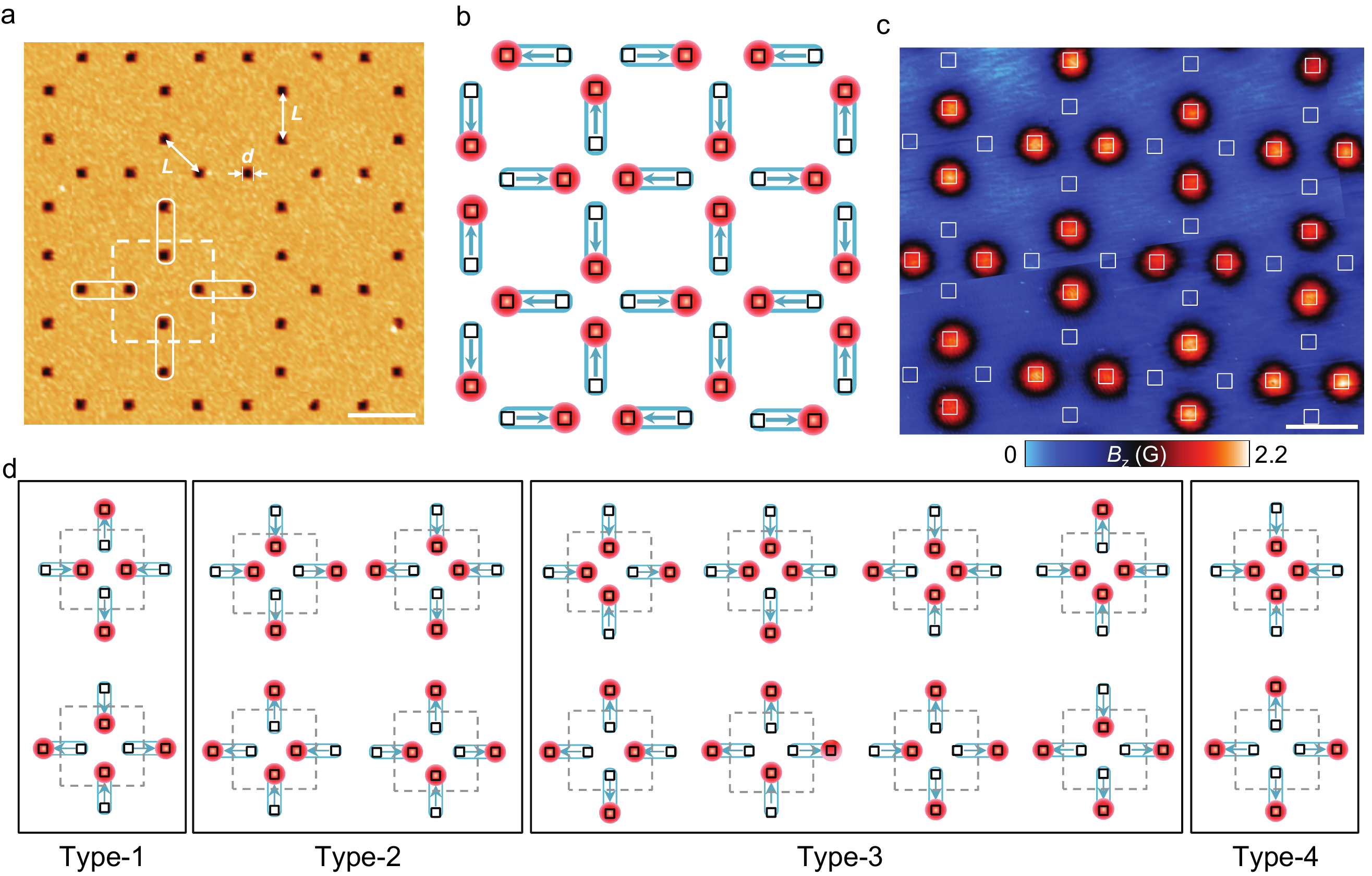}
\caption{(Color online) Vortex ice in a nanostructured superconductor. (a) Atomic-force microscope image of the sample. The dashed square indicates one unit cell. (b) Schematic presentation of the ground state ordering of the vortex lattice. Two antidots are paired to imitate an elongated double-well pinning site. The arrows in each pair, pointing from the empty antidot to the occupied one, represent the corresponding ground state arrangement of artificial spin ice. (c) Scanning Hall probe microscope image, showing the magnetic field distribution of the ground state vortex ice, observed after field-cooling at half matching field. The squares indicate the positions of antidots. (d) Possible vortex configurations corresponding to the magnetic-moment configurations of an artificial square spin ice, grouped by increasing energy (type-1 to type-4). The scale bars in (a) and (c) equal 4~$\mu$m.}
\label{fig1}
\end{figure*}

Vortices in type-II superconductors are known to exhibit strong, long-range, electromagnetic interactions. Each vortex carries the same quantized magnetic flux $\Phi_0=h/2e$ ($h$, Planck constant; $e$, elementary charge). It is energetically favorable for vortices to sit on artificial pinning centers (e.g., antidots) where superconductivity is locally suppressed \cite{Moshchalkov,Ge-PRB}. Compared with ASI, the density and size of vortices as well as the vortex-vortex/vortex-pin interaction strength can easily be tuned by changing temperature and/or magnetic field. This makes the vortex system  a perfect platform to study many-particle interactions on a potential-energy landscape with well controlled geometries \cite{Karapetrov,Ray,Cun}. 
Libal and coworkers proposed that, by arranging double-well pinning sites into square and kagome lattices, vortex ice states can be realized \cite{Libal}. Anomalous matching effects have been revealed in such vortex ice states by transport measurements \cite{Latimer,Trastoy}. The anomalous matching effect can be tuned by simply changing temperature to switch on/off the vortex ice state \cite{Trastoy}. So far, the direct observation of the vortex ice local configurations in real space is lacking.

Here we report the results of scanning Hall probe microscopy (SHPM) measurements on a nanostructured superconductor, which visualize and thus directly demonstrate the formation of vortex ice. We also analyze and classify different defects observed in the vortex-ice ground-state pattern. This analysis reveals new types of defects as compared to those observed earlier in ASI systems. We demonstrate the possibility to efficiently control the density and type of defects in the vortex ice configuration by changing the applied magnetic field. Note that such control can not be achieved in ASI systems.  

\section{2. Experimental}
The nanostructured pattern was fabricated with double layer of PMMA (300 nm) by electron beam lithography using a Raith e-beam system on a SiO$_2$/Si substrate . 
A superconducting Pb film of 90 nm was then electron-beam evaporated in ultrahigh vacuum ($3\times10^{-8}$ Torr) with a rate of 1 $\textrm{\AA/s}$ on the pattern. The substrate was cooled to 77 K by liquid nitrogen to ensure the homogeneous growth of Pb. A 10 nm Ge capping layer is deposited on top of the Pb film to protect it from oxidation. Then the multi-layer system was lift off in acetone,   resulting in the nanostructured film with antidot lattice (holes in Ge/Pb film). After liftoff,  the nanostructured film was transferred to a sputtering machine, then it was covered with a 35 nm layer of Au which plays the role of conducting layer for the tunneling junction of the scanning tunneling microscope (STM) tip. $T_c=7.35$ K is determined from local ac susceptibility measurements. 

The sample was imaged by using a low-temperature scanning Hall probe microscope in a lift-off mode \cite{Ge-nano}. An STM tip is assembled together with the Hall cross to make a Hall probe. The Hall probe was first brought to close proximity of the sample surface by using an STM tip \cite{Supplementary}. Then the probe was lifted 300 nm. This ensures a non-invasive measurement of the vortex pattern. In all the measurements, the applied field is perpendicular to the surface of the sample.

\begin{figure*}[!t]
\centering
\includegraphics*[width=0.9\linewidth,angle=0]{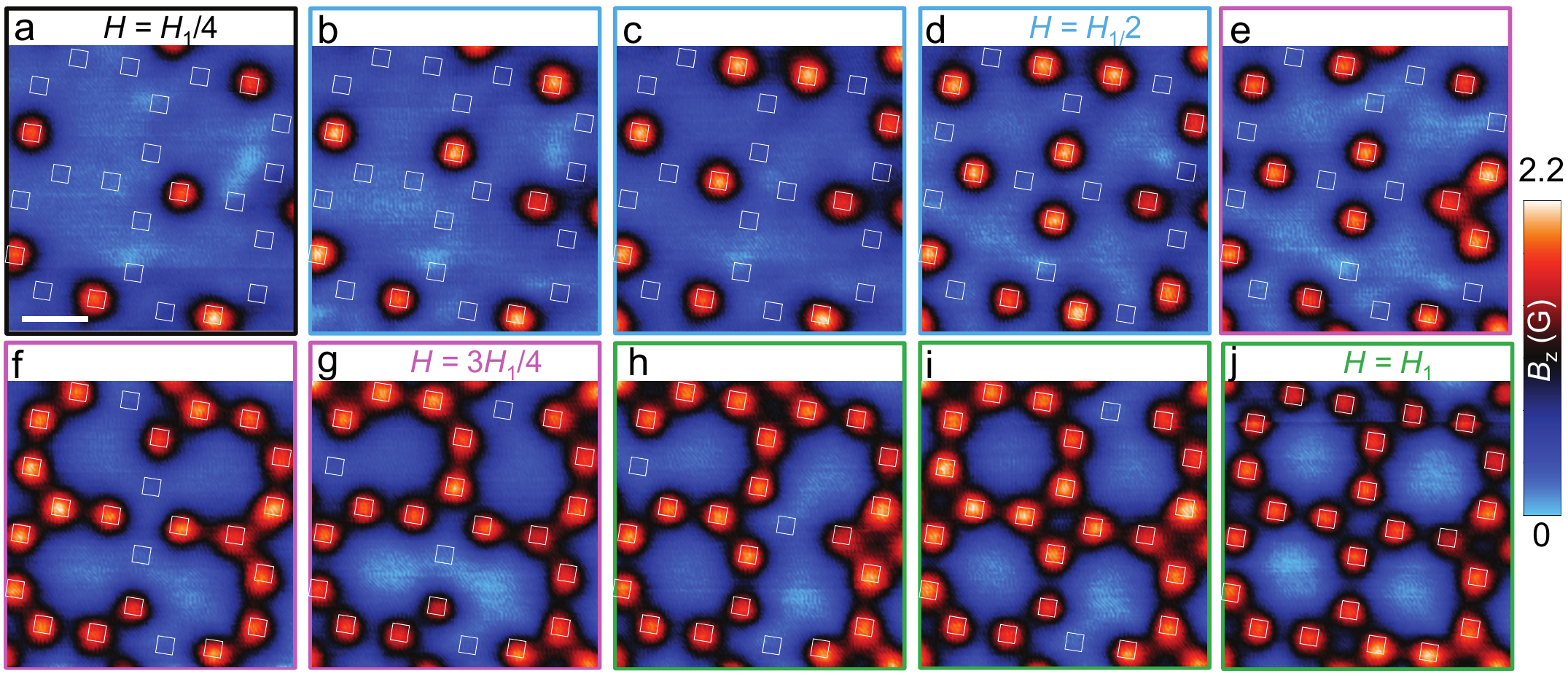}
\caption{(Color online) Vortex pattern evolution with magnetic field. Vortex images taken at $H=H_1/4=0.45$ Oe (\textbf{a}), $H=0.6$ Oe (\textbf{b}), $H=0.72$ Oe (\textbf{c}), $H=H_1/2=0.9$(\textbf{d}), $H=1.02$ Oe (\textbf{e}), $H=1.2$ Oe (\textbf{f}), $H=3H_1/4=1.35$ Oe (\textbf{g}), $H=1.65$ Oe (\textbf{h}), $H=1.7$ Oe (\textbf{i}), $H=H_1=1.8$ Oe (\textbf{j}). The white squares indicate the positions of antidots. At $H=H_1/4$, there is only one vortex in each vertex. Above $H_1/4$, two-vortex vertices appear and their number increases until a perfect ground state of vortex ice is formed at $H=H_1/2$. With increasing field, three-vortex vertices appear so that at $H=3H_1/4$ each vertex carry three vortices. At higher fields, four-vortex vertices are formed and eventually all the antidots are occupied by vortices at $H_1$. The images with the same frame color correspond to the maximum number of vortices per vertex equal to 1 (black), 2 (blue), 3 (red) and 4 (green). Scale bar equals 4~$\mu$m.}
\label{fig1}
\end{figure*}

\section{3. Results and Discussion}
As shown in Fig.~1a, the sample contains pairs of square holes arranged in such a way that four of those pairs meet at each vertex point of a square lattice. Each pair of holes imitates an elongated double-well pinning site proposed by Libal and coworkers \cite{Libal}. The array has the same distance $L=2.8~ \mu$m between the nearest antidot neighbors. The square holes have the size of $0.8\times0.8$~$\mu$m$^2$ in order to make the pinning potential deep enough to pin vortices. The sample geometry yields the first matching field of $H_1=1.8$~Oe, at which each antidot is occupied by one single-quantum vortex.  

$\\$
\textbf{3.1. Observation of the ground state vortex ice}\\

At half matching field, only one antidot is occupied by a vortex in each pinning pair. The geometry of the pinning array does not allow for all the vortices to be located at pinning sites with the same inter-vortex distance. This implies that the pairwise interactions cannot be simultaneously minimized, resulting in the geometrical frustration of the vortex lattice. To minimize the interaction energy, the ground state vortex arrangement is expected to follow the ice rule: two-in/two-out, i.e., two antidots are occupied at each vertex. We can attribute an arrow to each double-well pinning site, pointing from the free to the occupied well, to mimic the spin alignment. As shown schematically in Fig. 1b, a perfect ground state of square spin ice \cite{Wang,Zhang} is reproduced. This suggests the universal underlying ice rule for the two mentioned systems, ie, spin ice and vortex ice systems. We have scanned an area of $60\times70$ $\mu$m$^2$, and we observe the fraction of type-I vertices up to 58.3 percent. If the vortex distributions in individual antidot pairs were completely random, one would expect the fraction of type-I vertex to be as small as 12.5\%. The observed fraction clearly indicates that the ice rule determines the vortex distribution.  Figure 1c shows a scanned area of $22\times26$ $\mu$m$^2$,  where the vortex ice ground state is observed at half matching field, with each vertex containing two vortices arranged by the ice rule. This experimental observation confirms the proposed configuration of vortex ice \cite{Libal}. We would like to mention that different cooling rates (0.01 to 10 K/s) have been used when generating the vortex patterns. However, no clear dependence of vortex patterns on cooling rate is observed. This might arise from the fact that,  in superconductors with strong pinning centers (relatively large antidots), vortex patterns are formed and pinned at temperatures very close to $T_c$. We have seen that, when cooling down the sample, vortex patterns are formed and pinned already at temperature 0.99$T_c$.

\begin{figure*}[!t]
\centering
\includegraphics*[width=1\linewidth,angle=0]{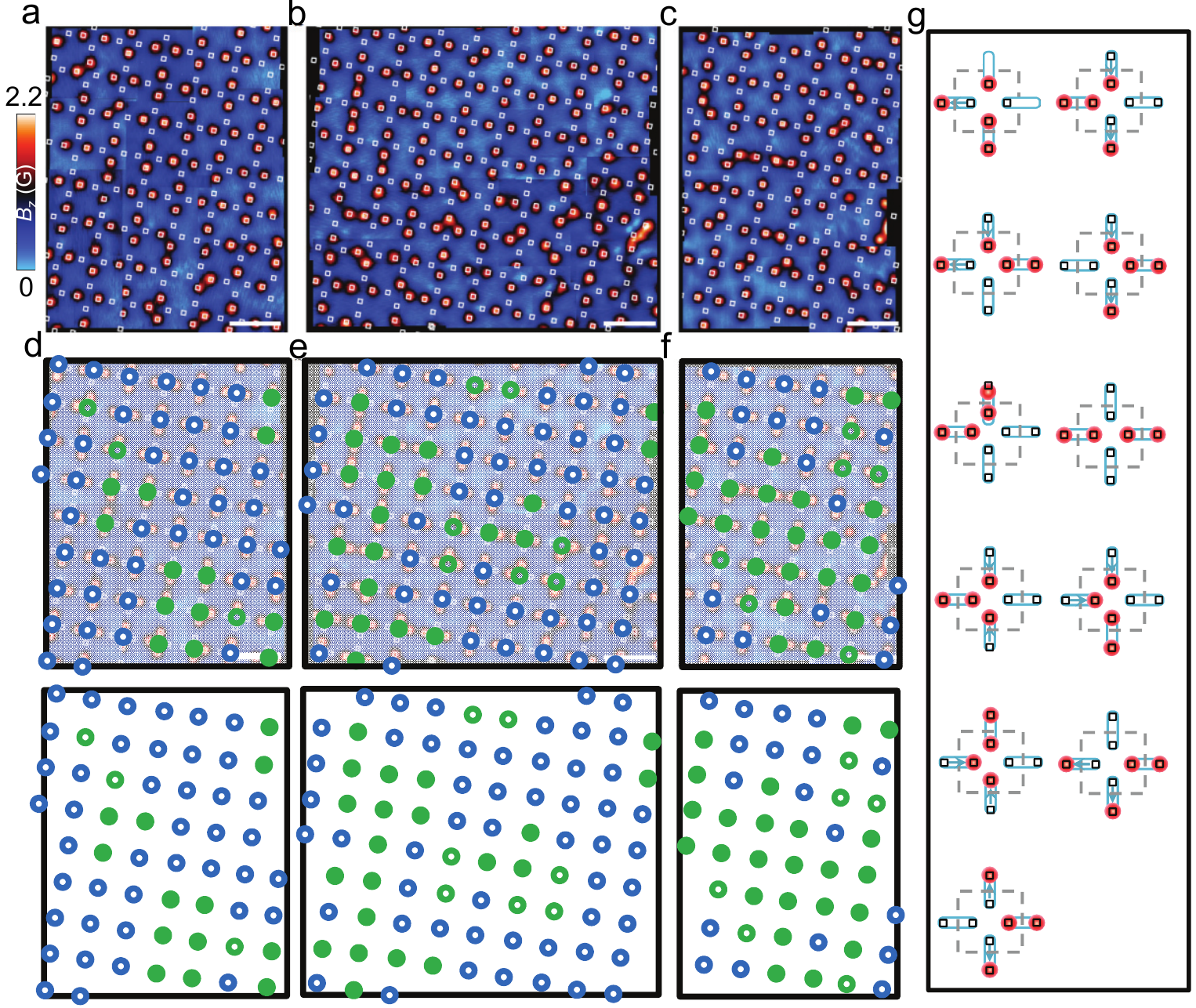}
\caption{(Color online) SHPM images showing the magnetic field distribution of the vortex ice pattern. Vortex patterns observed at  $H_1$/2-0.03 Oe (\textbf{a}), $H_1$/2 (\textbf{b}) and $H_1$/2+0.045 Oe (\textbf{c}). The white squares indicate the positions of antidots. \textbf{d-f} Vertex configurations corresponding to the SHPM images of \textbf{a-c}. The blue open circles represent vertices where the ice rule for the ground state is obeyed. The green open circles correspond to ice  defects of type-2 to type-4 shown in Fig. 1d. The solid circles indicate the non-ASI-like defects shown in \textbf{g}.}
\label{fig1}
\end{figure*}

In artificial square spin ice, besides the ground state, fourteen more possible vertex configurations can form with distinct symmetry or orientation \cite{Morgan}. According to the value of the interaction energy, all the configurations can be classified into four types.  The same vertex configurations are expected to appear in vortex ice as schematically shown in Fig.~1d (grouped with increasing energy). Both type-1 and type-2 configurations follow the ice rule, while type-3 and type-4 configurations can be considered as high energy defects in a vortex ice system. 
In ASI systems, long-range ordered ground state has never been observed due to the appearance of defects, which are manifested as monopole excitations \cite{Morgan}.  While in the ASI, considered in Ref. 14, the number of spins is fixed, the vortex density of vortex ice systems  can easily be tuned by varying magnetic fields. This allows us to study how different vertex configurations emerge and evolve from the ground state (at half matching field) in vortex ice systems. 

\begin{figure*}[!t]
\centering
\includegraphics*[width=0.85\linewidth,angle=0]{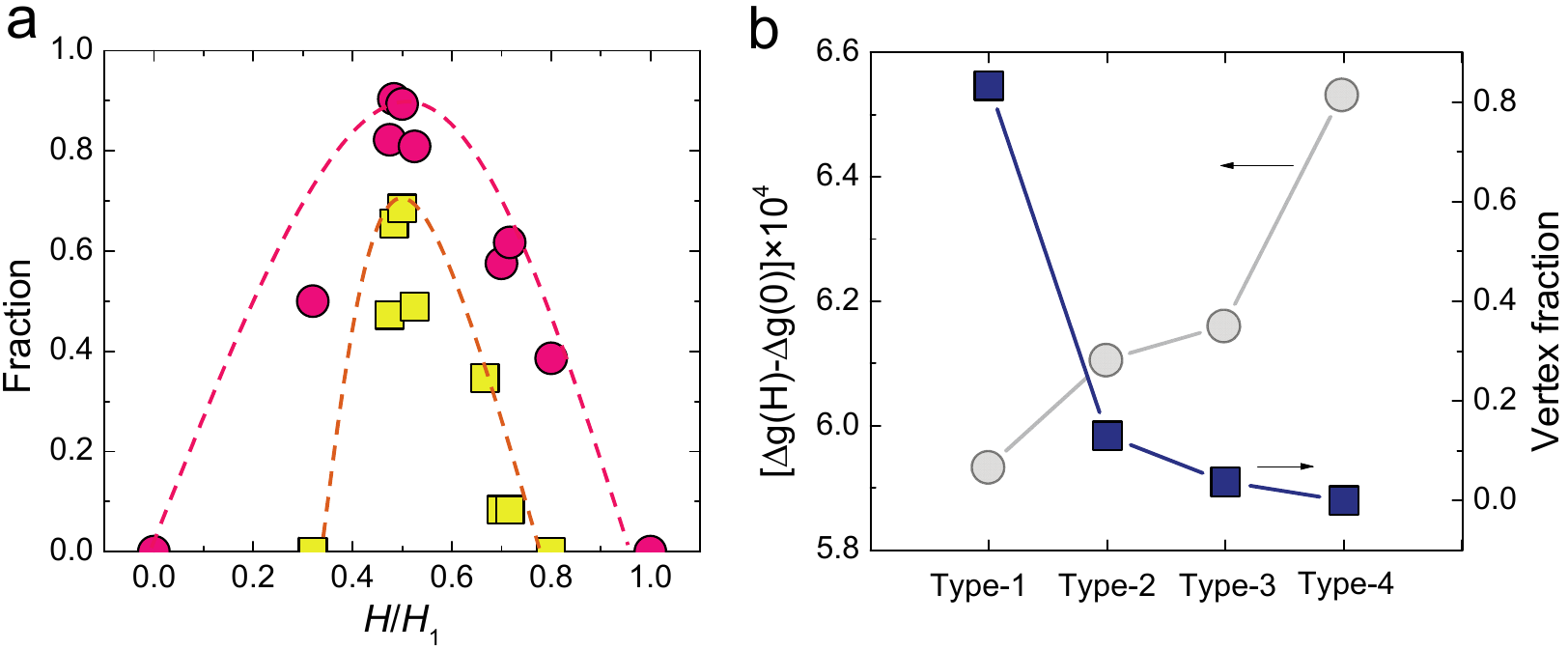}
\caption{(Color online) Statistics of vertex configurations in vortex ice. (\textbf{a}) Field dependence of the fraction of the spin-like pinning pairs (circles) and the vertices (squares) corresponding to the four type of vortex configurations in  Fig.~1d. Both the fraction of spin-like pinning pairs and ASI-like vertices reach maximum at half matching field. The dashed lines are guides to eyes. (\textbf{b}) Gibbs free energy density (circles) for different vertex configurations defined in Fig. 1d. The fraction of different ASI-like vertex configurations (squares) at half matching field. Type-4 vertex is not detected in the scanned area. Solid lines are guide to the eyes.}
\label{fig1}
\end{figure*}
$\\$
\textbf{3.2. Vortex ice state evolution with magnetic field}\\

Figure 2 presents the vortex pattern evolution with increasing magnetic field. A scanned area is chosen to observe the perfect ice ground state at half matching field. Each pattern is scanned after performing field cooling at the corresponding magnetic field value. When cooling down the sample, the vortex pattern freezes at a temperature very close to $T_c$ and it does not change with further decreasing temperature (see supplementary Figure s3). At $H=H_1/4$, each vertex accommodates one vortex. However, due to the pinning landscape geometry, it is not energetically favorable to put all the pinned vortices at the same distance from each other \cite{Trastoy}. At higher fields, two-, three- and four-vortex vertices start to form above $H_1/4$, $H_1/2$ and $3H_1/4$, respectively. The number of free antidots in the vortex pattern in Fig.~2g equals to that of occupied antidots in Fig.~2a. The frustration of the system is manifested by a high degeneracy of the vortex patterns. We have observed clear variations of vortex patterns in a series of field-cooling measurements at the same magnetic field (see supplementary Figure s4). We notice that no giant vortices are formed below the first matching field.  Above $H=H_1$, interstitial vortices start to appear (see supplementary Figure s5). 

In the dumbbell model of ASI, the ground state obeys the ice rule with the total magnetic charge $Q$ at each vertex being equal to zero \cite{Mengotti}. Type-3 defects, as excitations, lead to $Q=\pm2q$ with $q$ the elementary charge of each pole/antipole, while type-4 defects are not observed. In our sample,  vortices can also be considered as magnetic charges. We can observe the magnetic charge order at the vertices with $Q=\Phi_0$ (Fig.~2a), 2$\Phi_0$ (Fig.~2d), 3$\Phi_0$ (Fig.~2g) and 4$\Phi_0$ (Fig.~2j). As a result of the given magnetic charge at each vertex, the vortices tend to avoid the occupation of all the antidots in each square below the first matching field. This leads to vortex chains and tunnels of vortex free area at relatively high magnetic fields (supplementary Figure s6). By using SHPM vortex visualization, we have demonstrated the ability to control vertex defects in vortex ice systems.

Now we focus on the vortex configurations at half matching field, where the vortex ice ground state is formed. 
Figures 3a-c display the vortex ice state around half matching field at $T=4.2$~K. The vortex patterns are formed after performing a field-cooling from above $T_c$. The corresponding vertex maps are shown in Fig. 3d-e, where the vertices are classified into three types: ground state vertex (blue open circles), ASI-like defects (red open circles, also see Fig.~1d), and non-ASI-like defects (solid circles, Fig.~3g). 
For non-ASI-like defects, there is no corresponding vertex configuration in ASI systems, since the mapping with an arrow (as previously introduced in Fig. 1b) does not work for fully occupied/empty double-well pinning pair.

In Figs.~3a-c, only the short-range ordered ground state is seen. This is similar to the disordered magnetic-moment distribution observed in an as-grown ASI sample \cite{Morgan}, where the thermal and magnetic-field-induced excitations lead to the complex energy landscape associated with frustration. Besides the ground state, type-2 and type-3 vertex configurations can be identified whereas due to the relatively high interaction energy, no type-4 vertices are observed. In addition to the ASI-like defects, we also observe a large number of non-ASI-like defects which were not considered in the theory \cite{Libal}. In non-ASI-like defects, at least one pair of pinning sites is occupied by two vortices. We notice that the non-ASI-like defects tend to form clusters. 

The double-well pinning pairs themselves can constitute different vortex configurations. Three types of pinning pairs can be identified: 1) spin-like pinning pair with only one antidot occupied by a vortex; 2) pinning pair with two free antidots; 3) pinning pair with two occupied antidots. Unoccupied and doubly occupied pinning pairs can be considered as an analog of defects in a 2D spin array itself, e.g., like out-of-plane spins. At zero field, all pinning pairs are free, while at first matching field all pinning pairs are fully occupied. We have analyzed the statistics of different types of pinning pairs based on 81 vertices in the scanned area. In the field range between zero and the first matching field, the number of spin like pinning pairs increases and its fraction reaches maximum at half matching as shown in Fig.~4a (circles). We have also studied the statistics of the vertex configurations at different magnetic fields. Figure 4a (squares) displays the fraction of ASI-like vertices (squares) as a function of applied magnetic field. Clearly, the number of ASI-like vertices reaches its maximum at half matching field, although there are still a substantial amount of non-ASI-like defects.

$\\$
\textbf{3.3. Simulations}\\

To reveal the origin of disorder in the vortex state observed at half matching field, we have applied the Ginzburg-Landau theory to estimate the Gibbs free energy for different vortex configurations shown in Fig. 1b. The free energy is calculated for the corresponding stationary solutions of the time-dependent Ginzburg-Landau (TDGL) equations (see \cite{Silhanek,Ge-NC} for details of the used formalism and numerical approach). The TDGL simulations are performed for a periodic system with a supercell size of $13.6\times13.6$ $\mu$m$^2$ ($170\times170$ grid points) and the same geometry and distriution of antidots as that in the experimental sample. The periodic boundary conditions \cite{Doria} are used at the supercell boundaries, while the standard condition for the superconductor/vacuum interface is applied at the boundaries of antidots. The necessary configuration of pinned vortices is determined by choosing an appropriate initial guess for the order parameter distribution in the TDGL simulations. In the calculations, the coherence length and the penetration depth of the superconducting film coincide with those of the experimental sample at $T=4.25$ K. 

In Fig. 4b we plot the quantity $\Delta g(H_1/2)-\Delta g(0)$, where $\Delta g(H_1/2)$ [$ \Delta g(0)$] is the difference in the Gibbs free energy density between a vortex state at half matching field [the vortex-free state at zero field] and the normal state at the same magnetic field and temperature. The energy density is expressed in units of $\mu_0H_{c2}/(2 \kappa^2)$, where $\mu_0$ is the vacuum magnetic permeability, while $H_{c2}$ and $\kappa$ are the second critical field of the superconductor and its Ginzburg-Landau parameter, respectively. In these units, one has $\Delta g(0)=-1$. 
As follows from Fig.~4b, the energy difference between the analyzed vortex configurations is quite small with our designed geometry parameters. This implies that even relatively small local inhomogeneities of the superconducting film or dispersion of the pinning-center parameters may be sufficient for the formation of higher energy vertex types. One way to avoid such distortions is to use a smaller pinning lattice constant $L$ and a lower potential barrier between two pinning centers in a pair \cite{Libal}.
As further seen from Fig. 4b, the observed number of vertices of different type is in a qualitative agrement with the expectations based on the corresponding energy values. Type-4 vertices, characterized by a relatively high energy as compared to the other three vertex types, are not detected in the scanned area. Among all the ASI-like vertices, the fraction of the vertex configurations, corresponding to the ground state, reaches 83 \%, which is much larger than the fraction of ground state vertices in an as-grown ASI sample \cite{Wang}. This suggests certain advantage of using vortex system to study ice-like states.

\section{4. Conclusion}
In conclusion, we have performed SHPM experiments to directly visualize vortex patterns in a superconducting film with a square lattice of paired antidots. At half matching field, the vortex ice ground state is observed which follows the ice rule: two vortices located close to the vertex and two vortices located far from the vertex. Unlike the artificial spin ice systems with fixed number of interacting particles (magnetic dipoles), the superconducting systems provide the possibility to investigate vortex ice formation at different vortex densities (magnetic field). We demonstrate that the number and type of defects can be well controlled by varying magnetic field. We have found new types of defects, which were not present in the artificial spin ice systems studied earlier. Our direct imaging of the vortex ice state clearly shows the feasibility of studying geometrical frustrations in superconducting systems. This also opens new possibilities for studying superconductors with other types of geometrical frustration, such as kagome and brick lattices.

$\\$
\textbf{Acknowledgements}\\

We thank the support from the Methusalem funding by the Flemish government, the Flemish Science Foundation (FWO) and the MP1201 COST action. J.T. also acknowledges support from the Research Council of Antwerp University (BOF) and from the Flemish Science Foundation (FWO) grant nr. G.0429.15N.

%$\\$
%\textbf{Author contributions}\\
%J.Y.G. and V.Z. made the sample. J.Y.G. performed the SHPM measurements. V.G., J.T. and J.T.D. did the simulations. J.Y.G., V.G. and V.V.M. wrote the manuscript. All authors contributed to the discussion and analysis of the results. $\\$


\begin{references}

\bibitem{Wannier} G. H. Wannier, \textit{Phys. Rev.}, \textit{79}, 357 (1950).
\bibitem{Harris} M. J. Harris, S. T. Bramwell, D. F. McMorrow, T. Zeiske, K. W. Godfrey, \textit{Phys. Rev. Lett.}, \textit{79}, 2554-2557 (1997).
\bibitem{Siddharthan} R. Siddharthan,  B. S. Shastry, A. P. Ramirez, A. Hayashi, R. J. Cava, S. Rosenkranz, \textit{Phys. Rev. Lett.}, \textit{83}, 1854-1857 (1999).
\bibitem{Ramirez} A. P. Ramirez, A. Hayashi, R. J. Cava, R. Siddharthan, B. S. Shastry,\textit{Nature}, \textit{399}, 333-335 (1999).
\bibitem{Bramwell} S. T. Bramwell, M. J. P. Gingras,  \textit{Science}, \textit{294}, 1495-1501 (2001).
\bibitem{Petit} S. Petit, E. Lhotel, B. Canals,  M. Ciomaga Hatnean,  J. Ollivier, H. Mutka, E. Ressouche,  A. R. Wildes, M. R. Lees, G. Balakrishnan,  \textit{Nat. Phys.}, \textit{12}, 746-750 (2016).
\bibitem{Pauling} L. Pauling,  \textit{J. Am. Chem. Soc.}, \textit{57}, 2680-2684 (1935).
\bibitem{Lau} G. C. Lau, R. S. Freitas, B. G. Ueland, B. D. Muegge, E. L. Duncan, P. Schiffer, R. J. Cava, \textit{Nat. Phys.}, \textit{2}, 249-253 (2006).
\bibitem{Pomaranski} D. Pomaranski, L. R. Yaraskavitch, S. Meng, K. A. Ross, H. M. L. Noad,	H. A. Dabkowska, B. D. Gaulin, J. B. Kycia, \textit{Nat. Phys.}, \textit{9}, 353-356 (2013).
\bibitem{Perrin} Y. Perrin, B. Canals, N. Rougemaille, \textit{Nature}, \textit{540}, 410-413 (2016).
\bibitem{Castelnovo} C. Castelnovo, , R. Moessner, S. L. Sondhi, \textit{Nature}, \textit{451}, 42-45 (2008).
\bibitem{Morris}  D. J. P. Morris, D. A. Tennant, S. A. Grigera, B. Klemke, C. Castelnovo, R. Moessner, C. Czternasty, M. Meissner, K. C. Rule, J.-U. Hoffmann, K. Kiefer, S. Gerischer, D. Slobinsky, R. S. Perry, \textit{Science}, \textit{326}, 411-414 (2009).
\bibitem{Ladak} S. Ladak, D. E. Read,  G. K. Perkins, L. F. Cohen, W. R. Branford, \textit{Nat. Phys.}, \textit{6}, 359-363 (2010).
\bibitem{Morgan} J. P. Morgan, A. Stein, S. Langridge,  C. H. Marrow, \textit{Nat. Phys.}, \textit{7}, 75-79 (2011).
\bibitem{Tokiwa} Y. Tokiwa, T. Yamashita, M. Udagawa, S. Kittaka, T. Sakakibara, D. Terazawa, Y. Shimoyama,
T. Terashima, Y. Yasui, T. Shibauchi, Y. Matsuda, \textit{Nat. Commun.}, \textit{7}, 10807 (2016).
\bibitem{Vedmedenko} E. Y. Vedmedenko, \textit{Phys. Rev. Lett.}, \textit{116}, 077202 (2016).
\bibitem{Heyderman} L. J. Heyderman, \textit{Nat. Nanotechnol.}, \textit{8}, 705-706 (2013).
\bibitem{Ortiz} A. Ortiz-Ambriz,  P. Tierno, \textit{Nat. Commun.}, \textit{7}, 10575 (2015).
\bibitem{Wang2016} Y.-L. Wang, Z.-L. Xiao, A. Snezhko, J. Xu, L. E. Ocola, R. Divan, J. E. Pearson, G. W. Crabtree, W.-K. Kwork, \textit{Science}, \textit{352}, 962-966 (2016).
\bibitem{Wang} R. F. Wang, C. Nisoli, R. S. Freitas, J. Li, W. McConville, B. J. Cooley, M. S. Lund, N. Samarth, C. Leighton, V. H. Crespi, P. Schiffer, \textit{Nature}, \textit{439}, 303-306 (2006).
\bibitem{Moller} G. Moller, R. Moessner, \textit{Phys. Rev. Lett.}, \textit{96}, 237202 (2006).
\bibitem{Qi} Y. Qi, T. Brintlinger, J. Cumings, \textit{Phys. Rev. B}, \textit{77}, 094418 (2008).
\bibitem{Zhang} S. Zhang, I. Gilbert, C. Nisoli, G.-W. Chern, M. J. Erickson, L. OBrien, C. Leighton, P. E. Lammert, V. H. Crespi, P. Schiffer, \textit{Nature}, \textit{500}, 553-557 (2013).
\bibitem{Farhan} A. Farhan, P. M. Derlet, A. Kleibert,	A. Balan, R. V. Chopdekar, M. Wyss, L. Anghinolfi, F. Nolting, L. J. Heyderman, \textit{Nat. Phys.}, \textit{9}, 375-382 (2013).
\bibitem{Gilbert} I. Gilbert, G.-W. Chern, S. Zhang, L. OBrien, B. Fore, C. Nisoli, P. Schiffer, \textit{Nat. Phys.}, \textit{10}, 670-675 (2014).
\bibitem{Gilbert-2} I. Gilbert, Y. Lao, I. Carrasquillo, L. OBrien, J. D. Watts, M. Manno, C. Leighton, A. Scholl, C. Nisoli, P. Schiffer, \textit{Nat. Phys.}, \textit{12,} 162-165 (2016).
\bibitem{Zhou} X. Zhou, G.-L. Chua, N. Singh, A. O. Adeyeye, \textit{Adv. Funct. Mater.}, \textit{26}, 1437-1444 (2016)
\bibitem{Ke} X. Ke, J. Li, C. Nisoli, P. E. Lammert, W. McConville, R. F. Wang, V. H. Crespi, P. Schiffer, \textit{Phys. Rev. Lett.}, \textit{101}, 037205 (2008).
\bibitem{Li} J. Li, X. Ke, S. Zhang, D. Garand, C. Nisoli, P. Lammert, V. H. Crespi, P. Schiffer,\textit{Phys. Rev. B.}, \textit{81}, 092406 (2010).
\bibitem{Budrikis}Z.  Budrikis, J. P. Morgan, J. Akerman, A. Stein, P. Politi, S. Langridge, C. H. Marrows, R. L. Stamps, \textit{Phys. Rev. Lett.}, \textit{109}, 037203 (2012).
\bibitem{Moshchalkov} C. C. de Souza Silva, J. Van de Vondel, M. Morelle, V. V. Moshchalkov, \textit{Nature}, \textit{440}, 651-654 (2006).
\bibitem{Ge-PRB} J.-Y. Ge, V. N. Gladilin, Cun Xue, Jacques Tempere, Jozef T. Devreese, Joris Van de Vondel, Youhe Zhou, Victor V. Moshchalkov, \textit{Phys. Rev. B}, \textit{93}, 224502 (2016).
\bibitem{Karapetrov} G. Karapetrov, J. Fedor, M. Iavarone, D. Rosenmann, W. K. Kwok, \textit{Phys. Rev. Lett.}, \textit{95}, 167002 (2005).
\bibitem{Ray} D. Ray, C. J. Olson Reichhardt, B. Janko, C. Reichhardt, \textit{Phys. Rev. Lett.}, \textit{110}, 267001 (2013).
\bibitem{Cun} C. Xue, J.-Y. Ge, A. He, V. S. Zharinov, V. V. Moshchalkov, Y. H. Zhou, A. V. Silhanek, J. Van de Vondel, \textit{Phys. Rev. B}, \textit{96}, 024510 (2017).
\bibitem{Libal} A. Libal, C. J. Olson Reichhardt, C. Reichhardt, \textit{Phys. Rev. Lett.}, \textit{102}, 237004 (2009).
\bibitem{Latimer} M. L. Latimer, G. R. Berdiyorov, Z. L. Xiao, F. M. Peeters, W. K. Kwok,\textit{Phys. Rev. Lett.}, \textit{111}, 067001 (2013).
\bibitem{Trastoy} J. Trastoy, M. Malnou, C. Ulysse, R. Bernard, N. Bergeal, G. Faini, J. Lesueur, J. Briatico, J. E. Villegas, \textit{Nat. Nanotech.}, \textit{9}, 710-715 (2014).
\bibitem{Ge-nano} J.-Y. Ge, V. N. Gladilin, J. Tempere, C. Xue, J. T. Devreese, J. Van de Vondel, Y. Zhou, V. V. Moshchalkov, \textit{Nat. Commun.}, \textit{7}, 13880 (2016).
\bibitem{Supplementary} See Supplemental Material at [URL will be inserted by publisher] for the description of used experimental technique.
\bibitem{Mengotti} E. Mengotti, L. J. Heyderman, A. F. Rodriguez, F. Nolting, R. V. Hügli, H. B. Braun, \textit{Nat. Phys.}, \textit{7}, 68-74 (2011).
\bibitem{Silhanek} A. V. Silhanek, V. N. Gladilin, J. Van de Vondel, B. Raes, G. W. Ataklti, W. Gillijns, J. Tempere, J. T. Devreese, V. V. Moshchalkov, Supercond. Sci. Techno.  \textbf{24}, 024007 (2011).
\bibitem{Ge-NC} Ge, J. Gutireeze, J.-Y., Gladilin, J. T. Devreese, V. V. Moshchalkov, \textit{Nat. Commun.} \textbf{6}, 6573 (2015).
\bibitem{Doria} M. M. Doria, J. E. Gubernatis, D. Rainer, Phys. Rev. B  \textbf{39}, 9573 (1989).


\end{references}
\end{document}